\documentclass[10pt,aps,onecolumn,superscriptaddress,notitlepage, twocolumn]{revtex4-2}
\usepackage{amssymb}
\usepackage[utf8]{inputenc}
\usepackage{amsmath}
\usepackage{eucal}
\usepackage{footmisc}
\usepackage{epsfig}
\usepackage{bm}
\usepackage{xcolor}
\usepackage{epsfig}
\usepackage{epstopdf}
\usepackage{subfigure}
\usepackage{graphicx}
\usepackage[section]{placeins}
\usepackage[toc]{appendix}
\usepackage{amsmath}
\usepackage{amsmath,graphicx,amssymb,braket,xcolor,subfigure,upgreek}
\usepackage[colorlinks, linkcolor=blue, citecolor=blue, urlcolor=blue, breaklinks=true]{hyperref}
\usepackage{microtype}
\usepackage{bbm}
\usepackage{color}
\usepackage{braket}
\usepackage{amsmath,amssymb,amsfonts,latexsym,color,dcolumn,bm,amsbsy}

\def\be{\begin{equation}}
\def\ee{\end{equation}}
\def\bea{\begin{eqnarray}}
\def\eea{\end{eqnarray}}

\begin{document}

\title{Strong photon antibunching effect in a double cavity optomechanical system with intracavity squeezed light}

\author{M. Amazioug}
\email{amazioug@gmail.com}
\affiliation{LPTHE, Department of Physics, Faculty of sciences, Ibn Zohr University, Agadir, Morocco}
\author{M. Daoud}
\email{m\_daoud@hotmail.com}
\affiliation{Department of Physics, Faculty of Sciences, University Ibn Tofail, Kénitra, Morocco}
\affiliation{Abdus Salam International Centre for Theoretical Physics,
Strada Costiera, 11, 34151 Trieste, Italy}
\author{S. K. Singh}
\email{singhshailendra3@gmail.com}
\affiliation{Graphene and Advanced 2D Materials Research Group
(GAMRG), School of Engineering and Technology, Sunway
University, Jalan Universiti, Bandar Sunway, Petaling Jaya,
47500, Selangor, Malaysia}
\author{M. Asjad}
\email{asjad\_qau@yahoo.com}
\affiliation{Department of Applied Mathematics and Sciences, Khalifa University, Abu Dhabi 127788, UAE}

\begin{abstract}
We study the behaviour of the second-order correlation function in a double cavity optomechanical system and a degenerate optical parametric amplifier (OPA) is placed in each cavity. The first cavity is additionally driven by a weak classical laser field. The occurrence of strong photon antibunching effect in these two coupled cavities is observed. For suitable values of optomechanical coupling strength as well as photon hopping process, the system can exhibit a very strong photon antibunching effect. Our study also shows that the unconventional photon blockade occurs in both coupling, i.e. the weak coupling as well as in the strong coupling regimes as compared to the conventional photon blockade which occurs only in the strong coupling regime. We get a very strong photon antibunching effect under the unconventional photon blockade mechanism than the conventional photon blockade mechanism. Our study can be also used for the generation of single photon in coupled nonlinear optomechanical systems.
\end{abstract}
\maketitle

\section{Introduction}
Cavity optomechanics is a rapidly developing research area for exploring the radiation-pressure-mediated interaction between optical and mechanical degrees of freedom \cite{Aspee, Xio}. Due to the nonlinear coupling in between the optical mode to mechanical oscillations, cavity optomechanical system offers a robust platform for studying many interesting quantum phenomena such as quantum entangled states \cite{asjad1, mamazioug2020PLA, mamazioug2020QIP,asjad2, mamazioug2020EPJD,asjadop, singh2021entanglementLG, mamazioug2018EPJD,berihu, asjad3}, weak force sensing \cite{huang2017robust, Nade}, squeezing \cite{collett1985squeezing, asjad4, Saif, asjad5}, precision measurements \cite{wang2018precision}, quantum teleportation \cite{asjad6, asjad7}, optomechanical induced transparency \cite{weis2010optomechanically, asjade1, saif22, singh2022tunable} and optomechanical induced absorption \cite{Pei, Kenan, asjade2}, ground state cooling of macroscopic objects \cite{wilson, asjadc1, david1, asjadc2}, photon and phonon blockade \cite{Rabbl, Noori, Saif1, Saif2}. Photon blockade (PB) is a nonlinear optical effect that suppresses completely the multiple-photon occupancy in a quantum mode and favours only the single photon state \cite{AImamoglu1997}.  The photon blockade mechanism also generate sub-Poissonian  light when the system is driven by a classical light field. So far,  the photon blockade effect is  also one of the major experimental schemes to generate on demand single-photon sources and hence  plays a significant role in present day quantum information technologies \cite{EKnill2001, PKok2007}.

Photon blockade phenomenon has been studied theoretically in various nonlinear quantum optical systems, e.g., Kerr-type nonlinear cavity  \cite{JQLiao2010, SGhosh2019}, cavity optomechanical systems \cite{Rabbl,HWang2015,GLZhu2018,FZou2019} whereas the experimental works investigated  the trapped atom-cavity system \cite{KMBirnbaum2005} and a photonic crystal cavity coupled to a single quantum dot \cite{AFaraon2008,AReinhard2012,KMuller2015}.  Moreover, we have two wells- known mechanisms to realize a strong PB effect in any quantum system. The first one is $(i)$ unconventional photon blockade (UCPB) \cite{TCHLiew2010,MBamba2011,MBajcsy2013} which is generated due to the destructive quantum interference between different quantum transition paths from the ground state to a two-excited state, and the second one is $(ii)$ conventional photon blockade (CPB) \cite{AImamoglu1997,WLeonski1994,LTian1992} which depends on the larger nonlinearities to change the energy-level structure of the system. Both UCPB and CPB mechanisms have been implemented experimentally  in  \cite{KMBirnbaum2005,AFaraon2008, AJHoffman2011, HJSnijders2018, CVaneph2018}  as well as theoretically studied  in \cite{SShen2019, DYWang2020, FZou2020, HZShenPRA2015, HFlayac2017, SJLiu2020, HZShen2015, YHZhou2016}.\\
Motivated by the above-mentioned works, 
in this paper, we propose a scheme to generate a strong photon antibunching effect in a double cavity optomechanical system. Each cavity also contains a degenerate optical parametric amplifier (OPA). These two cavities are also spatially separated and coupled through the  single photon hopping process. We investigate photon antibunching effect in both the weak and  the strong coupling regimes through studying the second-order correlation function in this system. Here, the first cavity is also driven by a weak classical laser field as shown in Fig \ref{schema}. We also discuss the occurrence of the   both types of photon blockade effect, i.e. unconventional photon blockade and conventional photon blockade  as well as impact of various physical parameters for achieving a strong photon antibunching effect.\\
This paper is organised as follows. In Section 2, we have introduced the  model Hamiltonian of our proposed optomechanical system. In Section 3, we have obtained  the analytical and numerical results
related  to the equal-time second-order correlation function to discuss PB effect. In this same Section,  we also give the optimal parameter values to achieve the strong photon antibunching effect and we discuss the CPB and the UCPB mechanisms in both weak and strong coupling regimes. We conclude our results in Section 4.

\section{The Model Hamiltonian}

\begin{figure}[!htb]
\includegraphics[width=0.99\columnwidth]{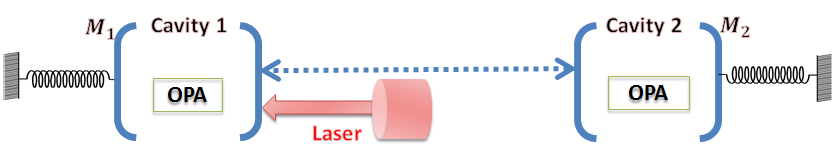}
\caption{Schematic of a double cavity optomechanical system.}
\label{schema}
\end{figure}

We consider an optomechanical system consisting of two Fabry-Pérot cavities coupled via the single photon hopping process (Fig \ref{schema}). Each cavity $j (j=1, 2)$ has a movable end-mirror $M_{j}$ and also contain inside a degenerate optical parametric amplifier (OPA) \cite{CSHu2017}. In our scheme, first cavity is also driven by a weak classical laser field with amplitude $\mathcal{E}$ and frequency $\omega_L$. The mass and the frequency of the $j^{th}$ movable mirror $M_{j}(j=1,2)$ are respectively denoted by $m_{j}$ and $\omega_{M_{j}}$. The mirror $M_j$ is coupled to the photons inside the cavity $j$ via radiation pressure. The coupling rate is $g_{j}=\frac{\omega_{a_{j}}}{L_j}\sqrt{\frac{\hbar}{m_j\omega{_M}_j}}$ with $L_j$ denoting the length of the $j{th}$ cavity \cite{Aspee}. The annihilation and creation operators of the $j{th}$ cavity mode are denoted by $\hat{a}_j$ and $\hat{a}_j^{\dagger}$ with $[\hat{a}_j, \hat{a}_j^{\dagger}]=1$ ($j=1,2$). The laser-cavity detuning is $\Delta_j=\omega_{L}-\omega_{a_{j}}$ where $\omega_{a_{j}}$ is the resonant frequencies of the cavity mode $jth$ cavity. The coupling strength of the photon hopping process is denoted by $J$. The total Hamiltonian describing the system in the rotating frame approximation   is given by ($\hbar=1$) 
\begin{equation} \label{eq:1} 
\widehat{\mathcal{H}} = \widehat{\mathcal{H}}_0+ \widehat{\mathcal{H}}_{PH},
\end{equation}
where 
\begin{equation} 
\widehat{\mathcal{H}}_{PH}=J(\hat{a}_1^{\dagger}\hat{a}_2+ \hat{a}_2^{\dagger}\hat{a}_1) 
\end{equation}
and the Free Hamiltonian $\widehat{\mathcal{H}}_0$ is the sum of three terms given as 
\begin{equation} \label{eq:2} 
\widehat{\mathcal{H}}_0 = \widehat{\mathcal{H}}_{free}+ \widehat{\mathcal{H}}_{om}+ \widehat{\mathcal{H}}_{drive},
\end{equation}
where the hermitians operators $\mathcal{H}_{free}$, $\mathcal{H}_{om}$ and $\mathcal{H}_{drive}$ written as
\begin{eqnarray}
\widehat{\mathcal{H}}_{free}=\sum_{j=1}^{2}[-\Delta_j \hat{a}^{\dagger}_{j}\hat{a}_{j}+\omega_{M_{j}}\hat{b}^{\dagger}_{j}\hat{b}_{j}],\\ 
\widehat{\mathcal{H}}_{om}=-\sum_{j=1}^{2}g_{j}\hat{a}^{\dagger}_{j}\hat{a}_{j}(\hat{b}^{\dagger}_{j}+ \hat{b}_{j}),\\
\widehat{\mathcal{H}}_{drive}=\mathcal{E}e^{i\phi}\hat{a}^{\dagger}_{1}+\mathcal{E}e^{-\i\phi}\hat{a}_{1},
\end{eqnarray}
where $\mathcal{E}=\sqrt{\frac{2\kappa_1 \wp}{\hbar\omega{_L}}}$ and $\phi$ are respectively the amplitude and the phase of the input coherent laser field. The quantity $\wp$ denotes the drive pump power of the input field. The cavity decay rate of the $jth$ cavity is $\kappa_j$. The annihilation and creation operators of the phonons mode of the $j{th}$ movable mirrors are represented as $\hat{b}_j$ and $\hat{b}^{\dagger}_j$ and satisfy $[\hat{b}_j, \hat{b}_j^{\dagger}]=1$ ($j=1,2$). We recall that the tunable photon statistics in parametrically amplified photonic molecules were studied in Ref. \cite{SShen2019} without optomechanical interaction. 
In addition to optomechanical interaction we take into account the degenerate optical parametric amplifier. This coupling is described by 
\begin{equation}
\widehat{\mathcal{H}}_{OPA}=\sum_{j=1}^{2}\left(i\lambda_{j} (\hat{a}^{\dagger}_{j})^2e^{i\theta}-i\lambda_{j} \hat{a}_{j}^2e^{-i\theta}\right),
\end{equation}
where the nonlinear gain and the phase of the field driving the OPA inside the $jth$ cavity are respectively given by $\lambda_{j}$ and $\theta$. The total Hamiltonian is now given by the sum of the Hamiltonian $\widehat{\mathcal{H}}$ (Eq.(\ref{eq:1})) and the interaction term given by
\begin{equation} \label{eq:3} 
\widehat{\mathcal{H}}_{tot} = \widehat{\mathcal{H}}+ \widehat{\mathcal{H}}_{OPA}.
\end{equation}
The degenerate OPA is pumped by another laser field at frequency $2\omega_L$ \cite{CSHu2019, amaziougEPJD2021}.
Using the following unitary transformation $\widehat{\mathcal{U}}=\exp[\widehat{\mathcal{S}}]$ with $\widehat{\mathcal{S}}=\sum^2_{j=1}\frac{g_j}{\omega_{m_j}}\hat{a}^{\dagger}_{j}\hat{a}_{j}(\hat{b}^{\dagger}_{j}-\hat{b}_{j})$; ($j=1, 2$), the Hamiltonian $\widehat{\mathcal{H}}_{tot}$ transform under this transformation as $\widehat{\mathcal{U}}^{\dagger}\widehat{\mathcal{H}}_{tot}\widehat{\mathcal{U}}=\widehat{\mathcal{H}}_1$ 
To derive this equation, we have used the Baker-Campbell-Hausdorff formula together with the commutation values of the operators $\hat{a}^{\dagger}$, $\hat{a}$, $\hat{b}^{\dagger}$ and $\hat{b}$.
The unitary transformation $\widehat{\mathcal{U}}$ decouples the two mechanical and the optical modes in the special case of \textit{weak optomechanical coupling}, i.e. $g_j/\omega_{m_j}\ll 1$ ($j=1, 2$) \cite{DYWang2020}. In fact, as we are interested by the photon statistic in the system, the Hamiltonian can be written as  
\begin{eqnarray}
\widehat{\mathcal{H}}_1 &=&\sum_{j=1}^{2}[-\Delta_j \hat{a}^{\dagger}_{j}\hat{a}_{j}-\mu_j(\hat{a}^{\dagger}_{j}\hat{a}_{j})^2+i\lambda (\hat{a}^{\dagger}_{j})^2e^{i\theta}-i\lambda \hat{a}_{j}^2e^{-i\theta}]\nonumber \\
&+&\mathcal{E}e^{i\phi} \hat{a}^{\dagger}_{1}+\mathcal{E}e^{-i\phi}\hat{a}_{1}+J(\hat{a}_1^{\dagger}\hat{a}_2+\hat{a}_2^{\dagger}\hat{a}_1), \label{eq:4} 
\end{eqnarray} 
where $\mu_j=g_j^2/\omega_{m_j}$ is the Kerr-type nonlinear strength.

\section{Photon Statistics}
In this section, we focus on the analytical solution of the non-Hermitian Schr\"odinger equation as well as numerical solution using Master equation. In addition, we discuss the generation of a strong photon antibunching effect in the weak and the strong coupling regimes. The non-Hermitian Hamiltonian is directly written by adding phenomenologically the imaginary decay terms as 
\begin{equation} \label{eq:5} 
\widehat{\mathcal{H}}_2=\widehat{\mathcal{H}}_1-i\frac{\kappa_1}{2}\hat{a}^{\dagger}_{1}\hat{a}_{1}-i\frac{\kappa_2}{2}\hat{a}^{\dagger}_{2}\hat{a}_{2}.
\end{equation}
The analytical expression of the correlation function can be obtained by solving the Schr\"odinger equation in the weak driving condition ($\mathcal{E}\ll \kappa_1$), i.e., $i\frac{\partial |\psi (t)\rangle}{\partial t}=\mathcal{H}_2|\psi (t)\rangle$, where $|\psi (t)\rangle$ is the state of the system. The evolution space can be limited in the low-excitation subspace, up to 2, i.e., $\mathcal{B}=\{|n_1,n_2\rangle/n_1+n_2\leq 2 \}$. The state of the system with the bare-state bases are
\begin{equation} \label{eq:6} 
|\psi (t)\rangle=\sum_{n_1, n_2}^{n_1+n_2 \leq 2}\mathcal{C}_{n_1n_2} (t) |n_1, n_2\rangle,
\end{equation}
where $\mathcal{C}_{n_1n_2} (t)$ is the probability amplitude of the bare state $|n_1, n_2\rangle$ with $n_1$ represents the photons being in the cavity $(1)$ and $n_2$ represents the photons being in the cavity $(2)$. For two identical cavities, $\kappa_1=\kappa_2=\kappa$, $m_1=m_2=m$, $\omega_{m_1}=\omega_{m_2}=\omega_m$, $\omega_{a_1}=\omega_{a_2}=\omega_a$, $L_1=L_2=L$, $\lambda_1=\lambda_2=\lambda$, $\Delta_1=\Delta_2=\Delta$ and $g_1=g_2=g$. Under the weak driving condition, the probabilities amplitudes satisfy the relation $|\mathcal{C}_{00}|\simeq 1\gg |\mathcal{C}_{01}|,|\mathcal{C}_{10}|\gg |\mathcal{C}_{11}|,|\mathcal{C}_{02}|,|\mathcal{C}_{20}|$. To simplify our purpose we consider $\theta=\phi=0$. The schr\"odinger equation leads to the following set of linear differential equations for the probability amplitudes $\mathcal{C}_{n_1n_2} (t)$
\begin{equation} \label{eq:7} 
i\frac{\partial \mathcal{C}_{00}}{\partial t} = \mathcal{E}\mathcal{C}_{10}-i\lambda(\mathcal{C}_{02}+\mathcal{C}_{20}),
\end{equation}
\begin{equation} \label{eq:8} 
i\frac{\partial \mathcal{C}_{01}}{\partial t}= \mathcal{E} \mathcal{C}_{00}-\Lambda\mathcal{C}_{01} + \sqrt{2}\mathcal{E}\mathcal{C}_{02}-J\mathcal{C}_{10},
\end{equation}
\begin{equation} \label{eq:9} 
i\frac{\partial \mathcal{C}_{10}}{\partial t} = \mathcal{E} \mathcal{C}_{00}-\Lambda\mathcal{C}_{10} + \sqrt{2}\mathcal{E}\mathcal{C}_{20}-J\mathcal{C}_{01},
\end{equation}
\begin{equation} \label{eq:10} 
i\frac{\partial \mathcal{C}_{11}}{\partial t}= -2\Lambda\mathcal{C}_{11} + \mathcal{E}\mathcal{C}_{01} + \sqrt{2}J\mathcal{C}_{02}- \sqrt{2}J\mathcal{C}_{20},
\end{equation}
\begin{equation} \label{eq:11} 
i\frac{\partial \mathcal{C}_{02}}{\partial t}= -2\Gamma\mathcal{C}_{02} -\sqrt{2}J\mathcal{C}_{11}+\i\sqrt{2}\lambda\mathcal{C}_{00},
\end{equation}
\begin{equation} \label{eq:12} 
i\frac{\partial \mathcal{C}_{20}}{\partial t}= -2\Gamma\mathcal{C}_{20} + \sqrt{2}\mathcal{E}\mathcal{C}_{10}-\sqrt{2}J\mathcal{C}_{11}+i\sqrt{2}\lambda\mathcal{C}_{00},
\end{equation}
where $\Lambda=\Delta+i\frac{\kappa}{2}-\mu$ and $\Gamma=\Delta+i\frac{\kappa}{2}-2\mu$. Eqs.(\ref{eq:7})-(\ref{eq:12}) can be solved analytically to obtain the dynamical state. Moreover, the steady-state result can be obtained by solving $\partial \mathcal{C}_{n_1n_2}/\partial t=0$, which can be simplified using some appropriate approximations as for instance ignoring those higher-order terms in the case of weak coupling interaction. The coefficients associated with one-photon ($n_1+n_2=1$) state are given by
\begin{equation} \label{eq:13} 
\mathcal{C}_{01}=\frac{J\mathcal{E}}{\Lambda^2-J^2},
\end{equation}
\begin{equation} \label{eq:14} 
\mathcal{C}_{10}=\frac{\Lambda \mathcal{E}}{\Lambda^2-J^2},
\end{equation}
and the coefficients associated with two-photon ($n_1+n_2=2$) states write as
\begin{equation} \label{eq:15} 
\mathcal{C}_{11}=\frac{J\left(- \mathcal{E}^2\Gamma-2i J^2\lambda+ \mathcal{E}^2\Lambda+2i\lambda\Lambda^2 \right)}{2(\Gamma\Lambda-J^2)(\Lambda^2-J^2)},
\end{equation}
\begin{equation} \label{eq:16} 
\mathcal{C}_{02}=\frac{ J^2\mathcal{E}^2\Gamma+ J^2\mathcal{E}^2\Lambda-2i J^2\lambda\Gamma\Lambda+2i\lambda\Gamma\Lambda^3}{2\sqrt{2}\Gamma(\Gamma\Lambda-J^2)(\Lambda^2-J^2)},
\end{equation}
\begin{equation} \label{eq:17} 
\mathcal{C}_{20}=\frac{J^2\mathcal{E}^2\Gamma-J^2\mathcal{E}^2\Lambda-i 2 J^2\lambda\Gamma\Lambda+2\mathcal{E}^2\Gamma\Lambda^2+ i 2\lambda\Gamma\Lambda^3}{2\sqrt{2}\Gamma(\Gamma\Lambda-J^2)(\Lambda^2-J^2)}.
\end{equation}
The equal-time second-order correlation functions defined by $g_j^{(2)}(0)=\langle \hat{a}^{\dagger}_j\hat{a}^{\dagger}_j\hat{a}_j\hat{a}_j \rangle/\langle \hat{a}^{\dagger}_j\hat{a}_j \rangle^2$ ($j=1,2$). It describes the probability to observe two photons in the $jth$ cavity at the same time. Using the results obtained here, are gets 
\begin{equation} \label{eq:18}
g_1^{(2)}(0)=\frac{2|\mathcal{C}_{20}|^2}{\overline{n}^2_1},\quad\, \quad g_2^{(2)}(0)=\frac{2|\mathcal{C}_{02}|^2}{\overline{n}^2_2},
\end{equation}
where $\overline{n}_1=|\mathcal{C}_{10}|^2+|\mathcal{C}_{11}|^2+2|\mathcal{C}_{20}|^2 \simeq |\mathcal{C}_{10}|^2$ and $\overline{n}_2=|\mathcal{C}_{01}|^2+|\mathcal{C}_{11}|^2+2|\mathcal{C}_{02}|^2\simeq |\mathcal{C}_{01}|^2$ represent the average photon occupations. The condition $g_j^{(2)}(0)>1$ corresponds to the photon bunching effect whereas $g_j^{(2)}(0)< 1$ leads to the photon antibunching effect and is characterized by the sub-Poissonian photon statistics \cite{Rabbl, Noori, Saif1}. Furthermore, the strong photon antibunching effect holds in the cavity (1) when $g_1^{(2)}(0)=0$, i.e., $\mathcal{C}_{20}=0$. On the other hand, the strong photon antibunching effect can be realized in the cavity (2) when  $g_2^{(2)}(0)=0$, i.e., $\mathcal{C}_{02}=0$. The Master equation approach which allows us to numerically calculate has the following expression

\begin{equation} \label{eq:21} 
\frac{d \rho}{d t}=-i[\widehat{\mathcal{H}_1},\rho]+\widehat{\mathcal{L}}_{\hat{a}_j}(\rho),
\end{equation}
where $\widehat{\mathcal{L}}_{a_j}(\rho)=\frac{\kappa}{2}(2\hat{a}_j\rho \hat{a}^+_j-\hat{a}^{\dagger}_j\hat{a}_j\rho -\rho \hat{a}^{\dagger}_j \hat{a}_j)$ and $\widehat{\mathcal{H}_1}$ is the Hamiltonian given by Eq.(\ref{eq:4}).

\subsection{Weak Coupling Regime}

In this subsection, we investigate the evolution of the equal-time second-order correlation function $g_j^{(2)}(0)$ ($j=1; 2$). We discuss the realization of the strong photon antibunching effect in weak coupling regime ($J<\kappa$ and $g\ll \omega_{m}$). The achieved photon blockade effect belongs to the UCPB (destructive quantum interference). 

The optimal parameter pairs $\lambda$ (the gain) and $\Delta$ (the cavity laser detuning) can be obtained together with the other fixed parameters by solving the equation $g_j^{(2)}(0)=0$ ($j=1; 2$). Here we also employ the parameters value considered in Ref. \cite{DYWang2020} : $J=0.95\kappa$, $\omega_m=2\pi\times 75\times 10^6$ Hz, $g =0.042\omega_m$, $\mathcal{E}=0.02\kappa$ and $\kappa=2\pi\times 0.15\times 10^6$ Hz.

The solution in the cavity (1) is $\{\{ \Delta^{(1)}_{opt} = -0.73\times  10^{-4}\omega_m , \lambda^{(1)}_{opt} = 0.93\times 10^{-6}\omega_m \}, \{ \Delta^{(1)}_{opt} = 0.15\times  10^{-2}\omega_m , \lambda^{(1)}_{opt} =-0.20\times  10^{-6}\omega_m \},\{ \Delta^{(1)}_{opt} = 33\times  10^{-4}\omega_m ~\text{Hz} , \lambda^{(1)}_{opt} = 1.47\times  10^{-6}\omega_m \}\}$. Moreover, the real solution of the optimal parameter pairs value in the cavity (2) is  $\{\{ \Delta^{(2)}_{opt} = -7.4\times 10^{-4}\omega_m , \lambda^{(2)}_{opt} = 0.33\times 10^{-6}\omega_m \}, \{ \Delta^{(2)}_{opt} = 0.21\times 10^{-2}\omega_m , \lambda^{(2)}_{opt} = -0.80\times 10^{-6}\omega_m \},\{ \Delta^{(2)}_{opt} = 0.47\times 10^{-2}\omega_m ~\text{Hz} , \lambda^{(2)}_{opt} = 0.40\times 10^{-6}\omega_m \}\}$.

\begin{figure}[!htb]
\begin{center}
\includegraphics[width=.47\textwidth]{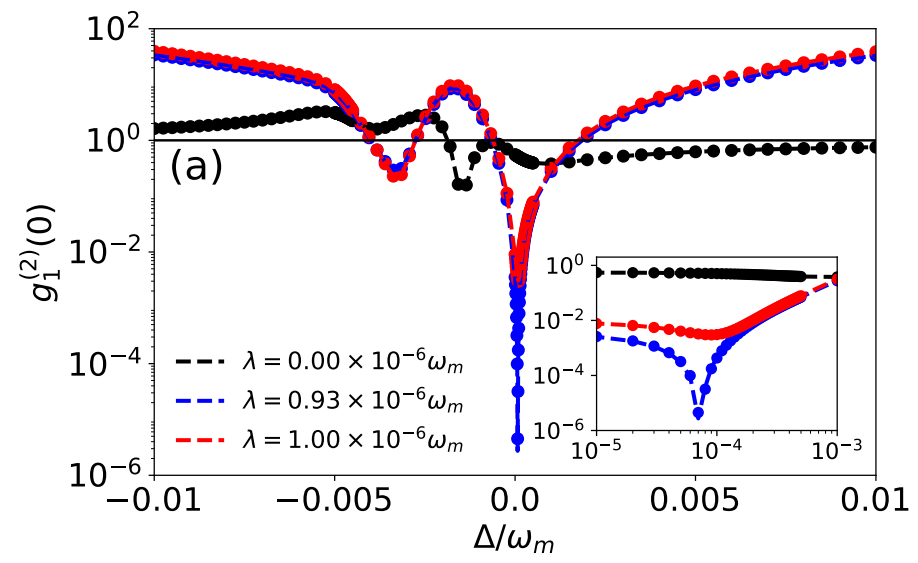}
\includegraphics[width=.47\textwidth]{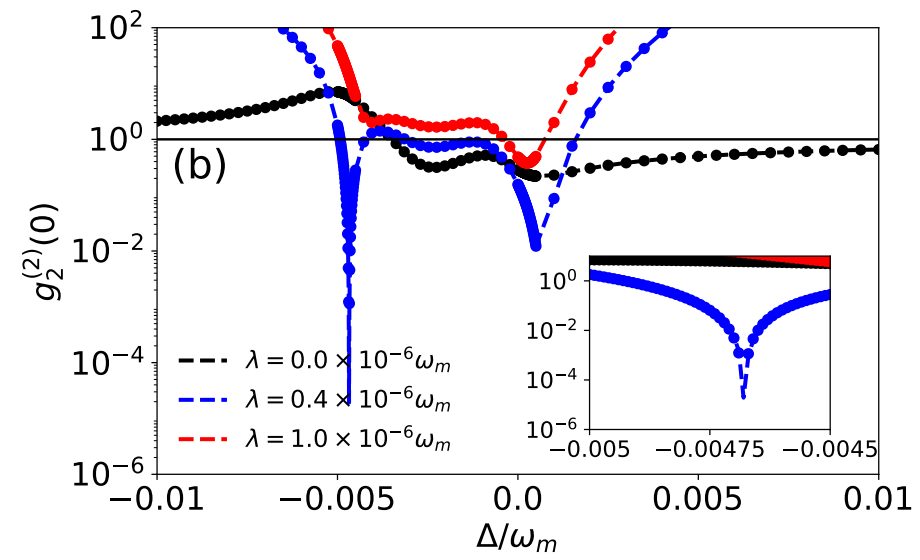}
\end{center}
\caption{Plot of the analytical (solid line) and numerical (dot) results in a double cavities of the equal-time second-order correlation function $g_(j)^{(2)}(0)$ ($j=1,2$) versus the cavity-laser detuning $\Delta/\omega_{m}$ (we use $\Delta \to -\Delta$ in this figure) for different values of the parameter $\lambda$ with $J=0.95\kappa$ and $g =0.042\omega_m$. $\lambda= 0.93\times 10^{-6} \omega_m$ with $\Delta=0.73\times 10^{-4}\omega_m$ in figure (a) in the cavity (1), and $\lambda=0.4\times 10^{-6} \omega_m$ with $\Delta=-0.47\times 10^{-2}\omega_m$ in figure (b) in the cavity (2). The solid horizontal line delimits the region under which represents the photon antibunching effect in the cavity.}
\label{PBWC}
\end{figure}
We plot in Fig \ref{PBWC} the equal-time second-order correlation function $g_1^{(2)}(0)$ as a function of the cavity-laser detuning $\Delta$ for different values of $\lambda$ by using analytic and numerical solutions. We note that the analytic and the numerical solution is obtained by solving master equation (Eq.\ref{eq:21}) are the same as shown in Figs \ref{PBWC} (a) and (b) which means that our results are correct. We remark that a strong photon antibunching effect occurs when we use the optimal values of parameter pairs $\{\Delta, \lambda\}$ listed above, i.e., $g_j^{(2)}(0)$ ($j=1, 2$) is much smaller than unity ($g_j^{(2)}(0)\ll 1$) at $\Delta^{(1)}_{opt}=-0.73\times10^{-4}\omega_m$  with $\lambda^{(1)}_{opt}= 0.93\times10^{-6}\omega_m$ in Fig \ref{PBWC} (a) and $\Delta^{(2)}_{opt}=0.47\times10^{-2}\omega_m$  with $\lambda^{(2)}_{opt}= 0.4\times10^{-6}\omega_m$ in Fig \ref{PBWC} (b). Moreover, when the parameter $\lambda$ does not take the optimal value $\lambda_{opt}$, the correlation function $g^{(2)}(0)$ cannot vanishes or tends to zero. We can explain the generation of a strong photon antibunching effect in the cavity (1) by the destructive quantum interference generated by different two photon between excitation schemes. The two-photon excited state $|0, 2 \rangle$ can be completely suppressed due to the ideal destructive quantum interference between in three different ways in \cite{SShen2019}: $(i)$ the direct excitation $|0, 0 \rangle \xrightarrow[]{\lambda} |2, 0 \rangle $ by a degenerate optical parametric amplifier interaction with the gain $\lambda$, $(ii)$ the direct excitation  
$|0, 0 \rangle  \xrightarrow[] {\mathcal{E}} |1, 0 \rangle \xrightarrow[]{\mathcal{E}} |2, 0 \rangle $
 by the driving field with the amplitude $\mathcal{E}$, and $(iii)$ the tunnelling-mediated transition $|1, 0 \rangle \overset{J}{\leftrightarrow} |0, 1 \rangle  
 \xrightarrow[]{\mathcal{E}}(|1,1 \rangle \xrightarrow[]{\text{J}} |0, 2\rangle)\xrightarrow{\text{J}}|2, 0\rangle $. 
From Fig \ref{PBWC}(b) the strong photon antibunching effects in the cavity (2) can be understand by the destructive quantum interference between directs the two paths : $(i)$ the direct excitation $|0, 0 \rangle \xrightarrow[]{\lambda} |0, 2 \rangle $ 
 by the degenerate optical parametric amplifier interaction with the gain $\lambda$, and $(ii)$ the tunnelling-mediated transition $|1, 0 \rangle \overset{J}{\leftrightarrow}  |0, 1 \rangle  \xrightarrow[]{\mathcal{E}} (|1,1 \rangle \xrightarrow[]{\text{J}}|2, 0\rangle )\xrightarrow[]{\text{J}}|0, 2\rangle $.

According to the above analysis and discussions, we find that the strong photon antibunching effect which occurs in the weak coupling regime ($J<\kappa$ and $g\ll \omega_m$) belongs to the class of unconventional photon blockade mechanism (the destructive quantum interference) \cite{DYWang2020}.\\
\begin{figure}[!htb]
\begin{center}
\includegraphics[width=.45\textwidth]{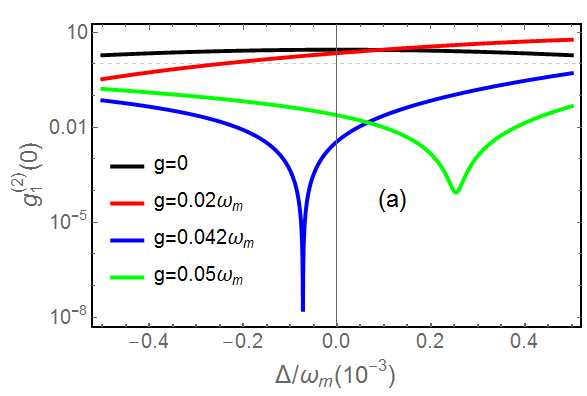}
\includegraphics[width=.45\textwidth]{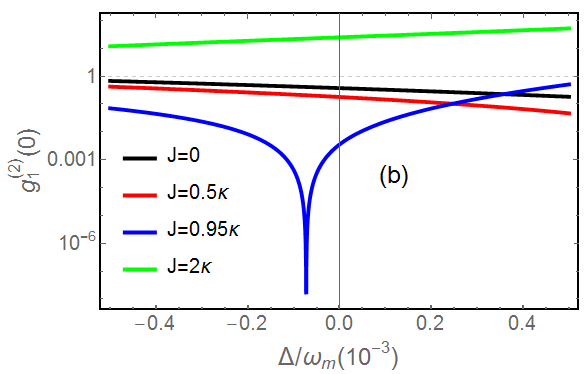}
\end{center}
\caption{(a) Plot of the equal-time second-order correlation function $g_1^{(2)}(0)$ versus the cavity-laser detuning $\Delta$ for various values of the optomechanical coupling strength $g$ with $\lambda^{(1)}_{opt}=0.93\times 10^{-6}\omega_m$ Hz and $J=0.95\kappa$. (b) Plot of the equal-time second-order correlation function $g_1^{(2)}(0)$ in the weak coupling region versus the cavity-laser detuning $\Delta$ for various values of the photon hopping coupling strength $J$ with $\lambda^{(1)}_{opt}=0.93\times 10^{-6}\omega_m$ and $g =0.042\omega_m$ Hz. The dashed horizontal line delimits the region under which represents the photon antibunching effect in the cavity (1).}
\label{WCVchiJ}
\end{figure} 
We plot in Fig \ref{WCVchiJ}(a) the equal-time second-order correlation function $g_1^{(2)}(0)$ versus the cavity-laser detuning $\Delta/\omega_m$ for various values of the optomechanical coupling strength $g$. This figure shows that $g_1^{(2)}(0)$ is very small when $g=0.042\omega_m$, i.e, strong photon antibunching effect occurs when $g=0.042\omega_m$. We remark that a strong photon antibunching effect is achieved for the exact value of $\Delta=-34304.2$ Hz as it is precisely predicted by the optimal parameters when $g=0.042\omega_m$ with $\lambda=437.053$ Hz. We notice that the equal-time second-order correlation function $g_1^{(2)}(0)$ is greater than 1 (the photon bunching effect) when $g=0$ with $\Delta \in [-400000~\text{Hz}, 300000~\text{Hz}]$. This means that the achieved photon antibunching effect is related to the optomechanical coupling $g$.\\
We plot in Fig \ref{WCVchiJ}(b) the equal-time second-order correlation function $g_1^{(2)}(0)$ versus the cavity-laser detuning for different values of the coupling $J$. This figure exhibits a strong photon antibunching effect for the value $\Delta^{(1)}_{opt}=1.54229\times 10^6$ Hz. This exactly the value expected via the parameter optimization when $J=0.95$ Hz with $\lambda^{(1)}_{opt}=692.709$ Hz. The equal-time second-order correlation function $g_1^{(2)}(0)$ is less than 1 when $J=0$ (the photon antibunching effect) for a wide range value of $\Delta$ as reported in Fig \ref{WCVchiJ}(b). This can be explained by the two-photon excitation path of the tunnelling-mediated transition, i.e., $|1, 0 \rangle \overset{J}{\leftrightarrow} |0, 1 \rangle  \xrightarrow[]{\mathcal{E}} |1,1 \rangle \overset{J}{\leftrightarrow}  |2, 0\rangle$. This means the coupling $J$ influences the occurrence of the photon antibunching effect.

\subsection{Strong Coupling Regime}

In this subsection, we will discuss both conventional and unconventional photon blockade mechanisms in the strong coupling regime. Moreover, the conventional photon blockade phenomenon would become obvious because the energy-level splitting induced by the strong coupling causes a large transition detuning between single-photon and two-photon states. Hence to discuss the conventional photon blockade in terms of the energy spectrum of different excitations, we consider the Hamiltonian of the system without the laser driving term, i.e., $\widehat{\mathcal{H}}_{ND} =\sum_{j=1}^{2}[-\Delta_j \hat{a}^{\dagger}_{j}\hat{a}_{j}-\mu_j(\hat{a}^{\dagger}_{j}\hat{a}_{j})^2+i\lambda (\hat{a}^{\dagger}_{j})^2 e^{i\theta}-i\lambda \hat{a}_{j}^2 e^{-i\theta}]+J(\hat{a}_1^{\dagger}\hat{a}_2+\hat{a}_2^{\dagger}\hat{a}_1)$. Eigenvalues of the single excitation $|0, 1\rangle$ and $|1, 0\rangle$ are $\epsilon_{1\pm}=J \pm \Delta \mp \mu$. The locations of CPB can be obtained using the theory of conventional blockade mechanism (resonant transition between the zero and single excitations) \cite{DYWang2020}
\begin{equation} \label{eq:19}
\Delta_{+}=\mu + J \quad ; \quad \Delta_{-}=\mu - J.
\end{equation}
Here, we consider the following values \cite{DYWang2020} : photon hopping coupling $J=8\kappa$, $\omega_m=2\pi\times 75\times 10^6$ Hz, $g =0.2\omega_m$, $\mathcal{E}=0.02\kappa$ and $\kappa=2\pi\times 0.15\times 10^6$ Hz. The parameter optimization of the pair of the parameter $\{\Delta, \lambda\}$ in the cavity (1) gives the following results $\{\{ \Delta^{(1)}_{opt} =  2.4\times 10^{-2}\omega_m , \lambda^{(1)}_{opt} = 1.1\times 10^{-6}\omega_m \}, \{ \Delta^{(1)}_{opt} = 4.0\times 10^{-2}\omega_m  , \lambda^{(1)}_{opt} = -0.8\times 10^{-6}\omega_m  \},\{ \Delta^{(1)}_{opt} = 6.0\times 10^{-2}\omega_m , \lambda^{(1)}_{opt} =  1.5\times 10^{-6}\omega_m \},\{ \Delta^{(1)}_{opt} =8.0 \times 10^{-2}\omega_m  , \lambda^{(1)}_{opt} = -0.12\times 10^{-6}\omega_m \},\{ \Delta^{(1)}_{opt} = 8.2\times 10^{-2}\omega_m , \lambda^{(1)}_{opt} = -0.02\times 10^{-6}\omega_m  \}\}$. Moreover, in the cavity (2) the optimal parameter pairs are $\{\{ \Delta^{(2)}_{opt} = 2.4\times 10^{-2}\omega_m  , \lambda^{(2)}_{opt} = 0.51\times 10^{-6}\omega_m  \}, \{ \Delta^{(2)}_{opt} = 4.0\times 10^{-2}\omega_m , \lambda^{(2)}_{opt} = -0.80\times 10^{-6}\omega_m  \},\{ \Delta^{(2)}_{opt} = 5.6\times 10^{-2}\omega_m  , \lambda^{(2)}_{opt} = 0.12\times 10^{-6}\omega_m  \},\{ \Delta^{(2)}_{opt} = 5.9\times 10^{-2}\omega_m  , \lambda^{(2)}_{opt} = 0.01\times 10^{-6}\omega_m  \}\}$.
\begin{figure}[!htb]
\begin{center}
\includegraphics[width=.47\textwidth]{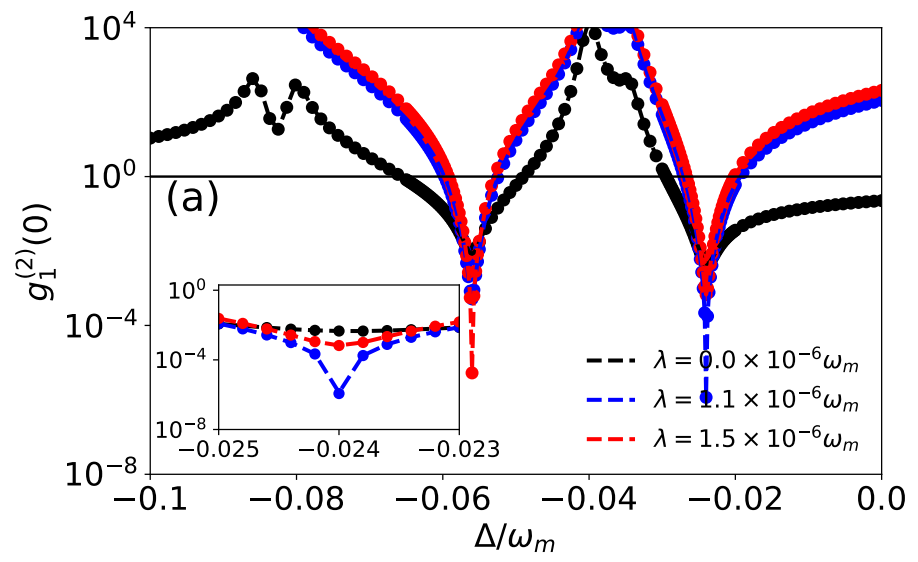}
\includegraphics[width=.47\textwidth]{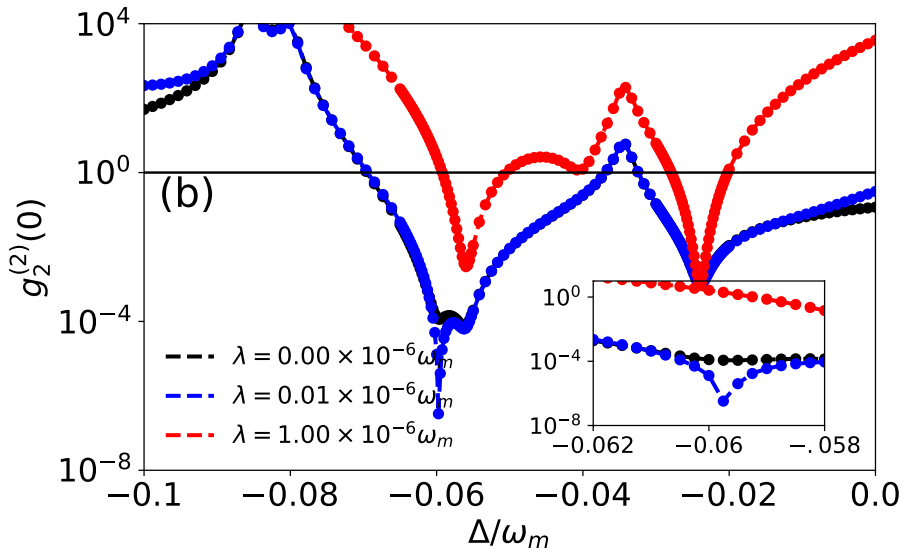}
\end{center}
\caption{Plot of the analytic (solid line) and numerical (dot) results in a double cavities of the equal-time second-order correlation function $g_j^{(2)}(0)$ ($j=1, 2$) in the strong coupling regime versus the cavity-laser detuning $\Delta/\omega_m$ (we use $\Delta \to -\Delta$ in this figure) for different values of the gain $\lambda$, (a) $\lambda^{(1)}_{opt}=1.1\times 10^{-6}\omega_m$ and (b) $\lambda^{(1)}_{opt}=0.01\times 10^{-6}\omega_m$, where $J=8\kappa$ and $g=0.2 \omega_m$. The solid horizontal line delimits the region under which represents the photon antibunching effect in the cavity (1) and (2).}
\label{CPB}
\end{figure}
We plot in Fig \ref{CPB} the equal-time second-order correlation function $g_j^{(2)}(0)$ ($j=1, 2$) versus the detuning $\Delta/\omega_m$ for various values of the gain $\lambda$. We also make a remark here that the analytical and the numerical solution is obtained by solving master equation (Eq.\ref{eq:21}) are the same as shown in this figure. The photon blockade locations are obtained from the Fig \ref{CPB}(a) for the values $\Delta=2.4\times 10^{-2}\omega_m$ and $\Delta=\Delta_+=\mu+J$. The small dips located is associated the CPB mechanism and the others dips are associated to the UCPB mechanism. We notice that the strong photon antibunching effect obtained under the UCPB mechanism is strong than one obtained under the CPB mechanism. We remark that a strong antibunching effect is realized when $g_1^{(2)}(0)$ takes small values ($g_j^{(2)}(0)\ll 1$) ($j=1, 2$). This can be explained by the unconventional photon blockade mechanism (the complete destructive quantum interference between different paths of two-photon excitation in the cavity (1) and (2)).

\begin{figure}[!htb]
\begin{center}
\includegraphics[width=.45\textwidth]{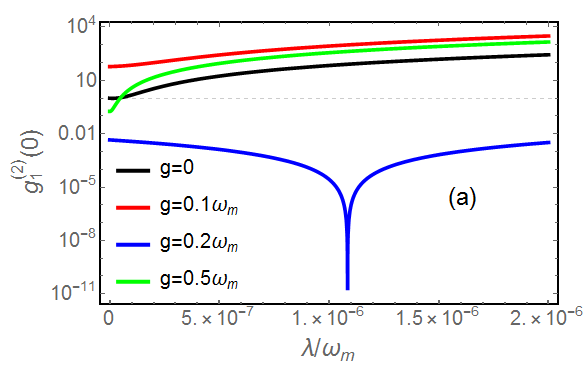}
\includegraphics[width=.45\textwidth]{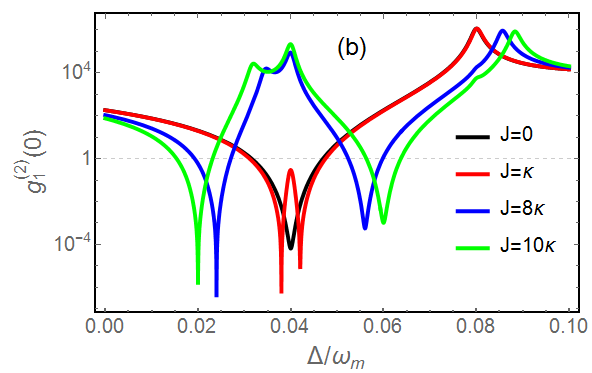}
\end{center}
\caption{(a) Plot of the equal-time second-order correlation function $g_1^{(2)}(0)$ versus the gain $\lambda$ for various values of the optomechanical coupling $g$ with $J=8\kappa$ and $\Delta=\Delta^{(1)}_{opt}=2.4\times 10^{-2}\omega_m$. (b) Plot of the equal-time second-order correlation function $g_1^{(2)}(0)$ versus the cavity-laser detuning $\Delta$ for various values of the photon hopping coupling strength $J$ with $\lambda^{(1)}_{opt}=1.1\times 10^{-6}\omega_m$ and $g=0.2\omega_m$. The dashed horizontal line delimits the region under which represents the photon antibunching effect in the cavity (1).}
\label{SCVchiJ}
\end{figure}
We plot in Fig \ref{SCVchiJ}(a) the equal-time second-order correlation function $g_1^{(2)}(0)$ versus the cavity-laser detuning $\Delta$ for various values of the optomechanical coupling $g$. This figure shows that $g_1^{(2)}(0)$ takes small values when $g=0.2\omega_m$ in this case a strong photon antibunching effect is achieved in the cavity (1). A strong photon blockade effect occurs for $\Delta=2.4\times 10^{-2}\omega_m$ as it is predicted by the parameters optimization when $g=0.2\omega_m$ with $\lambda=1.1\times 10^{-6}\omega_m$. We remark that when $g=0$ the equal-time second-order correlation function became greater than 1 ($g_1^{(2)}(0)>1$) i.e., the photon bunching effect for large values of $\lambda$. This means that the optomechanical coupling influences the generation of the photon antibunching.
We plot in Fig \ref{SCVchiJ}(b) the equal-time second-order correlation function $g^{(2)}(0)$ versus the cavity-laser detuning for different values of a photon hopping coupling $J$. We remark that a strong photon blockade effect occurs for various values of $J$ and $\Delta$ as shown in Fig \ref{SCVchiJ}(b). For example when $J=8\kappa$ a strong photon antibunching effect is achieved for $\Delta=2.4\times 10^{-2}\omega_m$. In agreement with the parameters optimization when $J=8\kappa$ with $\lambda=1.1\times 10^{-6}\omega_m$. And this shows the role of the photon hopping coupling $J$ in the generation of a strong photon antibunching effect. 

\section{Conclusion}

We have discussed a strong photon blockade effect through the analytical and numerical  evaluation of the second-order correlation function in a double-cavity optomechanical system. Here the two optical cavities are also coupled through the single photon hopping process. A degenerate optical parametric amplifier is placed inside each cavity. The first cavity is also driven by a weak classical laser field. We have obtained the conditions for  strong photon antibunching effect  in each optical cavity  for which the second-order correlation function satisfy $g^{(2)}(0)\ll 1$.  We have also analysed the photon blockade mechanisms in the weak as well as the strong coupling regimes for the conventional and unconventional photon blockade mechanisms. We have shown that the UCPB is achieved even for the weak coupling regime ($J<\kappa$ and $g\ll \omega_{m}$), whereas CPB occurs only in the strong coupling regime. We have discussed in details the effect of the optomechanical coupling strength and the photon hopping coupling on the generation of strong photon antibunching effects.  Our present study on the single-photon blockade and its generation in coupled optomechanical systems  can be of significant interest for the various applications in quantum information processing and quantum communication.


\begin{thebibliography}{2}
\bibitem{Aspee} M. Aspelmeyer, T.J. Kippenberg and F. Marquardt,  Rev. Mod. Phys. \textbf{86}, 1391 (2014).

\bibitem{Xio} H. Xiong, L.G. Si, Sci. China, Phys Mech Astron. \textbf{58}, 1 (2015).

\bibitem{asjad1} C. F Ockeloen-Korppi, E. Damskagg, J-M. Pirkkalainen, M. Asjad, A. A. Clerk, F. Massel, M.J. Woolley, M.A Sillanpaa,
Nature \textbf{556}, 478  (2018).

\bibitem{mamazioug2020PLA} M. Amazioug, B. Maroufi and M. Daoud, Phys. Lett. A \textbf{384}, 126705 (2020).

\bibitem{mamazioug2020QIP} M. Amazioug, B. Maroufi and M. Daoud,  Quantum Inf. Process. \textbf{19}, 16 (2020).

\bibitem{asjad2} J. Manninen, M. Asjad, E. Selenius, R. Ojajarvi, P. Kuusela, F. Massel, Physical Review A \textbf{98}, 043831 (2018).
\bibitem{mamazioug2020EPJD} M. Amazioug, B. Maroufi and M. Daoud,  Eur. Phys. Jour. D \textbf{74}, 9 (2020).

\bibitem{asjadop} M. Asjad, P. Tombesi, D. Vitali, Optics Express \textbf{23},  7786 (2015).

\bibitem{singh2021entanglementLG} S.K. Singh, J. Peng, M. Asjad and M. Mazaheri, J. Phys. B \textbf{54}, 215502,(2021).

\bibitem{mamazioug2018EPJD}M. Amazioug, M. Nassik and N. Habiballah, Eur. Phys. Jour. D \textbf{72}, 9 (2018).

\bibitem{berihu} B. Teklu, T. Byrnes, F. Khan, Phys. Rev. A 97 (2018) 023829.

\bibitem{asjad3} M. Asjad, M. A. Shahzad, F. Saif, The European Physical Journal D  \textbf{67}, 1 (2013).

\bibitem{huang2017robust} S. Huang and G. S. Agarwal, Phys. Rev. A \textbf{95} 023844, (2017).

\bibitem{Nade} A. Motazedifard, F. Bemani, M. Naderi, R. Roknizadeh, D. Vitali, New J. Phys. \textbf{18}, 073040(2016).
	
\bibitem{collett1985squeezing} M. J. Collett and D. F. Walls, Phys. Rev. A \textbf{32}, 2887, (1985).

\bibitem{asjad4}  M. Asjad, S. Zippilli, D. Vitali,  Phys. Rev. A \textbf{93}, 062307 (2016).

\bibitem{Saif} A. Kundu and S. K. Singh, Int. J. Theor. Phys. \textbf{58}, 2418 (2019).

\bibitem{asjad5} M. Asjad, G. S. Agarwal, M. S. Kim, P. Tombesi, G. Di. Giuseppe, D. Vitali, Phys. Rev. A \textbf{89}, 023849 (2014).
\bibitem{wang2018precision} Q. Wang, Laser Physics \textbf{28},075201 (2018).

\bibitem{asjad6} M. Asjad, S. Zippilli, P. Tombesi, D. Vitali,  Physica Scripta \textbf{90}, 074055 (2015).

\bibitem{asjad7} M. Asjad, P. Tombesi, D. Vitali, Phys. Rev. A \textbf{94}, 052312 (2016).

\bibitem{weis2010optomechanically} S. Weis, R. Rivilere, S. Delleglise, E. Gavartin, O. Arcizet, A. Schliesser and T. J. Kippenberg, Science \textbf{330}, 1520, (2010).

\bibitem{asjade1} M. Asjad, Journal of Russian Laser Research \textbf{34}, 159 (2013).

\bibitem{saif22} S.K. Singh, M. Parvez, T. Abbas, Jia-Xin Peng, M. Mazaheri, M. Asjad, Phys. Lett. A \textbf{442}, 128181 (2022).


\bibitem{singh2022tunable} S. K. Singh, M. Asjad, and C.H. Raymond Ooi, Quantum Inf. Process. \textbf{21},18, (2022).

\bibitem{Pei} X. Y. Zhang, Y. Q. Guo, P. Pei  and X. X. Yi, Phys. Rev. A \textbf{95}, 063825 (2017).

\bibitem{Kenan} Kenan Qu and G. S. Agarwal,  Phys. Rev. A \textbf{87}, 031802 (2013).

\bibitem{asjade2} M. Asjad, Journal of Russian Laser Research \textbf{34}, 278 (2013).

\bibitem{wilson} I. Wilson-Rae, N. Nooshi, W. Zwerger, and T. J. Kippenberg Phys. Rev. Lett. \textbf{99}, 093901 (20067).

\bibitem{asjadc1} M. Asjad, S. Zippilli, D. Vitali, Physical Review A \textbf{94}, 051801 (2016).

\bibitem{david1}  M. Rossi, N. Kralj, S. Zippilli, R. Natali, A. Borrielli, G. Pandraud, E. Serra, G. Di. Giuseppe, D. Vitali, Phys. Rev. Lett \textbf{119}, 123603 (2017).

\bibitem{asjadc2} M. Asjad, N. E. Abari, S. Zippilli, D. Vitali, Optics Express \textbf{27}, 32427 (2019).

\bibitem{Rabbl} P. Rabl, Phys. Rev. Lett. \textbf{107}, 063601 (2011).

\bibitem{Noori} J.-Q. Liao and F. Nori, Phys. Rev. A. \textbf{88}, 023853 (2013).

\bibitem{Saif1} S.K. Singh and  S. V. Muniandy, Int. J. Theor. Phys. \textbf{55}, 287 (2016).

\bibitem{Saif2}	S. K. Singh and C. H. Raymond Ooi, J. Opt. Soc. Am. B \textbf{31}, 2390, (2014).

\bibitem{AImamoglu1997} A. Imamoglu, H. Schmidt, G. Woods and M. Deutsch, Phys. Rev. Lett. \textbf{79}, 1467 (1997).

\bibitem{EKnill2001} E. Knill, R. Laflamme and G. J. Milburn, Nature (London) \textbf{409}, 46 (2001).
\bibitem{PKok2007} P. Kok, W. J. Munro, K. Nemoto, T. C. Ralph, J. P. Dowling and G. J. Milburn, Rev. Mod. Phys.\textbf{ 79}, 135 (2007).

\bibitem{JQLiao2010} J.-Q. Liao and C. K. Law. Phys. Rev. A. \textbf{82}, 053836 (2010).

\bibitem{SGhosh2019} S. Ghosh and T. C. H. Liew. Phys. Rev. Lett. \textbf{123}, 013602 (2019).

\bibitem{HWang2015} H. Wang, X. Gu, Y.-x. Liu, A. Miranowicz and F. Nori, Phys. Rev. A \textbf{92}, 033806 (2015).

\bibitem{GLZhu2018} G.-L. Zhu, X.-Y. L, L.-L. Wan, T.-S. Yin, Q. Bin and Y. Wu, Phys. Rev. A \textbf{97}, 033830 (2018).

\bibitem{FZou2019} F. Zou, L.-B. Fan, J.-F. Huang and J.-Q. Liao, Phys. Rev. A \textbf{99}, 043837 (2019).

\bibitem{KMBirnbaum2005} K. M. Birnbaum, A. Boca, R. Miller, A. D. Boozer, T. E. Northup and H. J. Kimble, Nature \textbf{436}, 87 (2005).

\bibitem{AFaraon2008} A. Faraon, I. Fushman, D. Englund D, N. Stoltz, P. Petroff and J. Vuckovic, Nat. Phys. \textbf{4}, 859 (2008).

\bibitem{AReinhard2012} A. Reinhard, T. Volz, M. Winger, A. Badolato, K. J. Hennessy, E. L. Hu and A. Imamoglu, Nat. Photonics \textbf{6}, 93 (2012).

\bibitem{KMuller2015} K. Muller, A. Rundquist, K. A. Fischer, T. Sarmiento, K. G. Lagoudakis, Y. A. Kelaita, C. Sanchez Munoz, E. del Valle, F. P. Laussy and J. Vuckovic, Phys. Rev. Lett. \textbf{114}, 233601 (2015).

\bibitem{TCHLiew2010} T. C. H. Liew and V. Savona, Phys. Rev. Lett. \textbf{104}, 183601 (2010).

\bibitem{MBamba2011} M. Bamba, A. Imamoglu, I. Carusotto and C. Ciuti, Phys. Rev. A. \textbf{83}, 021802(R) (2011).


\bibitem{MBajcsy2013} M. Bajcsy, A. Majumdar, A. Rundquist and J. Vuckovic, New J. Phys. \textbf{15}, 025014 (2013).

\bibitem{WLeonski1994} W. Leonski and R. Tanas, Phys. Rev. A \textbf{49}, R20 (1994).

\bibitem{LTian1992} L. Tian and H. J. Carmichael, Phys. Rev. A \textbf{46}, R6801 (1992).

\bibitem{AJHoffman2011} A. J. Hoffman, S. J. Srinivasan, S. Schmidt, L. Spietz, J. Aumentado, H. E. T\"ureci and A. A. Houck, Phys. Rev. Lett. \textbf{107}, 053602 (2011).

\bibitem{HJSnijders2018} H. J. Snijders, Phys. Rev. Lett. \textbf{121}, 043601 (2018).

\bibitem{CVaneph2018} C. Vaneph, A. Morvan, G. Aiello, M. F\'echant, M. Aprili, J. Gabelli and J. Esteve, Phys. Rev. Lett. \textbf{121} , 043602 (2018).

\bibitem{SShen2019} S. Shen, Y. Qu, J. Li and Y. Wu, Phys. Rev. A, 100(2), 023814 (2019).
\bibitem{DYWang2020} D. Y. Wang, C. H. Bai, S. Liu, S. Zhang and H. F. Wang, New J. Phys. \textbf{22}, 093006 (2020).

\bibitem{FZou2020} F. Zou, D. G. Lai and  J. Q. Liao, Opt. express \textbf{28}, 16175-16190 (2020).

\bibitem{HZShenPRA2015} H. Z. Shen, Y. H. Zhou and X. X. Yi, Phys. Rev. A \textbf{91}, 063808 (2015).

\bibitem{HFlayac2017} H. Flayac and V. Savona, Phys. Rev. A \textbf{96}, 053810 (2017).

\bibitem{SJLiu2020} J. S. Liu, J. Y. Yang, H. Y. Liu and A. D. Zhu, Opt. Express \textbf{28}, 18397-18406 (2020).

\bibitem{HZShen2015} H. Z. Shen, Y. H. Zhou, H. D. Liu, G. C. Wang and X. X. Yi, Opt. express \textbf{23}, 32835-32858 (2015).

\bibitem{YHZhou2016} Y. H. Zhou, H. Z. Shen, X. Q. Shao and X. X. Yi, Opt. Express \textbf{24}, 17332-17344 (2016).

\bibitem{CSHu2017} C. S. Hu, X. R. Huang, L. T. Shen, Z. B. Yang and H. Z. Wu, Eur. Phys. J. D \textbf{71}, 24 (2017).

\bibitem{CSHu2019} C. S. Hu, L. T. Shen, Z. B. Yang, H. Wu, Y. Li and S. B. Zheng, Phys. Rev. A \textbf{100}, 043824 (2019).

\bibitem{amaziougEPJD2021} M. Amazioug and M. Daoud, Eur. Phys. J. D \textbf{75}, 178 (2021).

\end{thebibliography}
\end{document}